\documentclass[15pt,a4paper]{article}
\usepackage{authblk}
\usepackage{graphicx}
\usepackage{float}
\usepackage{color}
\usepackage{xcolor}
\usepackage{caption}
\usepackage{subcaption}
\usepackage{makeidx}
\usepackage{array}
\usepackage[T1]{fontenc}
\usepackage[font=small,labelfont=bf]{caption}
\usepackage{graphicx}
\def\dbar{{\mathchar'26\mkern-12mu d}}
\newcolumntype{P}[1]{>{\centering\arraybackslash}p{#1}}
\newcolumntype{M}[1]{>{\centering\arraybackslash}m{#1}}
\author[]{K. Nilavarasi$^{a}$}
\author[]{M. Ponmurugan$^{b}$\footnote{ponphy@cutn.ac.in}}
\affil[]{$^{a}$School of Education and Training, Department of Education, \\
$^{b}$School of Basic and Applied Sciences, Department of Physics, \\
Central University of Tamil Nadu, Thiruvarur - 610 005, Tamil Nadu, India.}
\date{}
\usepackage[left=2cm,right=2cm,top=2cm,bottom=2cm]{geometry}

\begin{document}
\title{ Efficiency at maximum $\dot{\Omega}$ figure of merit of a spin half quantum heat engine
in the presence of external magnetic field}

\maketitle
\section*{Abstract} 
We consider a finite time quantum heat engine analogous to finite time classical Carnot heat engine with a working substance of spin half particles. We study the efficiency at maximum  $\dot{\Omega}$ figure of merit of the quantum heat engine of spin half particles as a working substance in the presence of external magnetic field. The efficiency of this engine at maximum $\dot{\Omega}$ figure of merit shows anomalous behavior in certain region of particles population levels. Further, we find that the efficiency at maximum $\dot{\Omega}$ figure exceeds all the known bounds and even approaches the Carnot efficiency at finite time. Our study indicates that the population of spin half particles plays a crucial role in quantum heat engine  whose collective effect in the quantum regime can provide superior engine performance with higher efficiency.   
   
\section{{Introduction}}
Quantum devices are going to outperform in the future technological era and hence a deep understanding of quantum phenomenon is much required. Thermodynamics, the physical theory,  was  developed primarily for  understanding and optimizing the large scale devices viz., heat engines \cite{Kondepudi}.  Despite this focused theory, thermodynamics includes many formulated universal laws explaining the classical thermodynamic phenomenons.  With the emerge of quantum technologies, the field of thermodynamics was extended to understand the processes in  the quantum regime.  The important aspect of quantum thermodynamics is to explore how quantum properties like  entanglement, superposition and coherence can be utilized in the optimization of performance of quantum devices \cite{Scully, zhang, dillen, Turkpence, hardal, huang, uzdin, Jaramillo, liu, thomas}.  Quantum heat engine is one such device utilises the quantum properties explores the influence of quantum effects on the thermodynamic behaviour of the system.  There are notable studies on the analysis of quantum thermal machine performance which proves that it will outperform its classical counterpart \cite{Scully, zhang, dillen, huang}.  \\

The ideal cycle  of heat engine, called Carnot cycle, explains the existence of maximum efficiency of heat engines.  The cycle  was given by Sadi Carnot and it consists of two reservoirs at two different temperature $T_h$ and $T_c < T_h$ and the heat engine working between these two reservoirs attain the maximum possible work and zero power.  This engine operates infinitely slow to achieve this maximum efficiency called Carnot efficiency $\eta_C=1-T_c/T_h$, which is an unrealistically high upper bound for practical devices operates  in finite time \cite{anderson, salamon}.  So the study of operation of heat engines at finite time becomes very important  and it was Yvon \cite{yvon},  Novikov \cite{novikov}, Chambadal \cite{chambadal} pioneered and later Curzon and Ahlborn \cite{curzon}, obtained the  efficiency at optimized power rather than the usual efficiency.  The efficiency at optimized maximum  power given by Curzon and Ahlborn is \cite{curzon}, 
\begin{equation}
\eta_{CA}=1-\sqrt{\frac{T_c}{T_h}} =\frac{\eta_c}{2}+ \frac{\eta_c^2}{8}+O(\eta_C).
\label{CA}
\end{equation}
This efficiency is found to provide a more realistic upper bound of efficiency for the engine  in endoreversible approximation \cite{curzon, callen} as well as under the condition of strong coupling between heat and work \cite{van, de, gomez}.  This finite time analysis becomes crucial in the quantum regime.  Hence the study of finite time process in quantum thermodynamics is inevitably required.  A list of literature analyzing quantum heat engines performance includes the works examining the influence  of  quantum correlations \cite{barrios}, many-body effects \cite{beau, li, yychen, venkatesh},  energy optimization \cite{singh}, degeneracy \cite{pena, albarian}, efficiency and power statistics \cite{denzler}, endoreversible cycles  \cite{sdeffner, pal}, finite-time cycles \cite{vcavina, hanggi}, comparisons between classical and quantum machines \cite{quan, garda, lutz}.  In particular, T  Feldmann  studied the four stroke Otto refrigerator  in a extremely short cycle times \cite{feldmann}.  Smith et al analysed endoreversible Otto engines \cite{sdeffner}.  It was also demonstrated that for the quantum Otto engine, the efficiency at maximum power depends on the equation of state  (nature of the medium).  The endoreversible quantum  Otto engine having working medium as ideal gases or harmonic oscillators are found to operate at the Curzon-Ahlborn efficiency \cite{sdeffner}.  However, there are also studies which shows that the complicated medium can exhibits efficiency significantly larger. Few works also showed that for a working medium of two non-interacting identical particles, the Otto quatum heat engine exhibited the enhanced performance for the particles of bosons and reduced performance for the fermions \cite{myer}.   On the other hand, Cakmark et.al., stated that irrespective of the working medium, the quantum Carnot cycle is reversible and it is also possible to extract useful work at the maximum Carnot efficiency, $\eta_C = 1-\frac{T_c}{T_h}$ \cite{selcuk}.  But here the problem in the construction of the reversible Quantum Carnot cycle is its scale invariance.  The simpler quantum systems inherently has the scale invariance whereas for  the complex systems it is difficult to establish the scale invariance.  When the working medium is not scale invariant, they may not be at thermal equilibrium which means when the working substance is brought in contact with the heat reservoir, even before the start of isothermal process, there exist an irrversible relaxation process  \cite{selcuk}.  These arguments naturally leads to a question that whether the efficiency of quantum engines in general is influenced by the optimal working medium?  \\

Other than medium, the choice of  the optimization parameter  also plays a vital role in identifing the performance limits of heat engines.  Recently, numerous reports analysed the performance of heat engines through various optimization process using finite time thermodynamics \cite{broeck,  sanchez, guo, medina}. The work by Esposito et.al., showed the bounds of efficiency for a low dissipation Carnot like heat engine under maximum power optimization process \cite{esposito}.  Ma  also found the per-unit time efficiency at the maximum power.  There are also other studies which also analysed the efficiency at maximum power conditions \cite{ma}.  The above studies are in the context of classical  finite time engines.  In the quantum regime, there are studies concerned  with the universal trade-off between power and efficiency of a quantum dot solar cell \cite{piet},  with the trade-off between geometric arguments namely, how the geometry of work fluctuations and effciency  are made in microscopic heat engines \cite{mehboudi, saito}.  Dorfmann studied the efficiency at maximum power of a laser quantum heat engine enhanced by noise induced coherence \cite{dorfman}.  Bera et. al considered the engines with  working systems consisting of  finite number of quantum particles and are restricted to one-shot measurements .  They showed that these engines in the one-shot finite size regime deliver maximum power with Carnot efficiency.  Two important points are to be considered here with all the refrenced studies above.  1.  All the reports stated above concerned with the efficiency of classical and quantum heat engines at optimized  power conditions and 2. similar to the classical engines, it is impossible for a quantum heat engine to yield maximum power at Carnot efficiency.   Although the efficiency at maximum power was as a desirable operational regime, there are also several other optimization parameters  which can enhance the performance of heat engines.  Two important parameters among them are namely, $\dot{\Omega}$ figure of merit and efficient power $\chi_{\eta P}$  \cite{omegaref, nila}.  The former, $\dot{\Omega}$  figure of merit, is a trade-off between the useful energy delivered and energy lost by heat engines and the later provides a compromise between efficiency $\eta$ and power, $P$.   These are respectively defined through $\dot{\Omega}= (2\eta- \eta_{max})\dot{Q_h}$ and $\chi_{\eta P}=\eta_{P}$, where, $\eta$ and $\eta_{max}$ are efficiency and maximum efficiency respectively and $\dot{Q_h}$ is the rate of heat flow in the hot reservoir per cycle time. These parameters are also found to enhance the heat engine performance and is getting much attention nowadays to study the heat engine performance \cite{nila}. \\ 

 In the present work,  we study the performance of two level quantum heat engine which has spin half particles in an external magnetic field as working medium by optimizing the $\dot{\Omega}$ figure of merit.  Thereby, the influence of spin half particles as medium can be investigated  and the extend to which the optimization parameters $\dot{\Omega}$ figure of merit enhance the performance of quantum heat engines can be addressed as well.  In the present study we will use the formalism of  quantum heat engine model developed by Y Bassie et.al.,  which is a two level quantum system modeled as a spin half particle in an external magnetic field \cite{bassie}.  The system is in contact with the two reservoirs: hot reservoir and cold reservoir with the inverse temperatures, $\beta_H$ and $\beta_C$ respectively. When the system is in contact with the hot reservoir, it absorbs heat and  perform some mechanical work.  It then releases the remaining heat to the cold reservoir which is at the inverse temperature $\beta_C$.  The expressions for work, heat and efficiency at each process of the cycle can be calculated.   We quantitatively compare the efficiency obtained under the considered optimizing parameters/figure of merit and elucidate the role of quantum system and its interaction  on the considered figures of merit.  \\

This paper is organized as follows: We begin by outlining the model, notation, and previous results that are of primarily important for our analysis and these contribute to section II of the manuscript.  In section III, the working of the model in the finite time cycles are explained.  Section IV discusses the optimization of efficiency of quantum heat engine at maximum $\dot{\Omega}$ figure of merit. This section also compares the extreme bounds of efficiency obtained from optimization of the above said parameter and their significance are discussed. The paper concludes with the conclusion in section V.\\

\section{The Model}
This section will elaborately explains the quantum thermodynamic cycles of a Carnot like heat engine model and the thermodynamic quantities like heat, work and efficiency are evaluated.  

We consider an irreversible quantum Carnot like heat engine for an arbitrary two level quantum system as working substance. The working substance under this study is the spin half particles in the presence of external time dependent magnetic field \cite{kos}. These spin half particles constitutes a two level system, as it can have two possible discrete energy states corresponding to the parallel and anti-parallel alignment of their spins  with respect to the applied field. The Hamiltonian of the above considered system can be written as, 
\begin{equation}
H=\frac{1}{2}\lambda \sigma_z
\end{equation}
where, $\lambda = \hbar \omega$, $\hbar$ is the Planck's constant, $\omega$ is the frequency and $\sigma_z$ is the Pauli spin matrix.  Here the value of $\lambda$ can be adjusted by simply changing the strength of the external magnetic field.   

The particles are assumed to be non-interacting and each are coupled to the heat baths $ B_1$ and $B_2$ with corresponding inverse temperatures $\beta_1$ and $\beta_2$  representing hot and cold reservoirs, respectively. The engine undergoes a Carnot like cycle having two  isothermal branches and two adiabatic branches.  Each of these branches are assumed be irreversible.   Figure \ref{f1} represent the illustration of the assumed heat engine model. The processes $A \rightarrow B$ and $C \rightarrow D$  are the quasi static (quantum) isothermal processes during which the working substance is always kept in thermal equilibrium with the heat bath. When the system is in contact with the thermal bath, there will be a change in populations of the discrete energy levels which is given by Boltzmann-Gibb’s distribution, 
\begin{equation}
P_{n}(i)= \frac{1}{Z_i} e^{-\beta_i E_n}.\\
\end{equation}
Here $n=1,2$ is the number of energy levels with energy $E_n$, $i$ represents the system state
in the cold and hot temperature reservoir, $Z_i= \sum_{n} exp(-\beta_i E_n)$  $(i=1,2)$ is the partition function and $\beta_1= (k_B T_c)^{-1}$ \& $\beta_2= (k_B T_h)^{-1}$ are inverse temperatures, $k_B$ is the Boltzmann constant.  The heat that is exchanged during the isothermal process can be obtained by, $dQ=T dS$, where $dS$ is the change in the entropy during this process. \\
The processes $B \rightarrow C$ and $D \rightarrow A$  are the quantum adiabatic processes during which the  working medium is detached from the reservoir.   During these processes there is no heat exchange between the working medium and the reservoirs.  There is also no change in the populations and hence $P_n(B) = P_n(C)$ and $P_n(D) = P_n(A)$.\\

When this system is placed in contact with the heat reservoir (cold reservoir) having inverse constant temperature $\beta_i$, the internal energy of the system is given by, 
\begin{equation}
U(i)= \langle E \rangle =\sum_{n} E_n(i) P_n(i).
\end{equation}
Here $P_n(i)$  is the probability of finding the system in the particular energy state $n(i)$ with energy $E_n(i)$ and $\langle . \rangle$ denotes average over the system energy states.  The change in the average energy of the system can occurs as a consequence of work and heat.  Hence, heat and work can also be extended to quantum-mechanical systems and  expressed as functions of the eigen energy values  and probability distributions.  The first law of thermodynamics can be generalized to quantum mechanical system and is given as follows:
\begin{eqnarray}
dU=d\langle E \rangle = \sum_{n} E_n dP_n +\sum_{n} P_n dE_n = \dbar Q + \dbar W,       
\label{internalenergy}
\end{eqnarray}
where $dU$ is the change in inetenal energy,
\begin{equation}
\dbar Q = \sum_{n} E_n dP_n
\end{equation}
and 
\begin{equation}
\dbar W = \sum_{n} P_n dE_n.
\end{equation}
Here $E_n$ is the $n^{th}$ eigen energy value corresponding to the eigenstate $|n \rangle$ of the quantum-mechanical system with the Hamiltonian $H$ under consideration and $ P_n$  is the occupation probability in the $n^{th}$ eigenstate.  According to the above equation~\ref{internalenergy}, the change in internal energy occurs in two situations: 1. When there is an infinitesimal change in the energy level $dE_n$ and 2. An infinitesimal change in the probabability $dP_n$.  The former occurs due to the change in external magnetic field at constant probability, due to which the work will be done by (on) the system  and the latter occurs when there is a heat exchange between the sysytem and the reservoir at constant field, thereby aborbing (or releasing) of heat by the system occurs.  The above two situations leads to the processes which are path dependent.  And hence, the $\dbar W$, the path dependent work done on(by) the system and the heat absorbed (released) by the spin half particle system  can be represented as, 
\begin{equation}
\dbar W=\sum_{n}P_n dE_n = P^{e} d\Delta
\label{workdone}
\end{equation}
and
 \begin{equation}
\dbar Q=\sum_{n}E_n dP_n = \Delta d P^{e}.
\end{equation} 
Here, $\Delta$ denotes the energy gap between the ground state and excited state and $P^{e}$ represents the probability of getting the spin half particle in the excited state of the system \cite{bassie}.  As we know from classical thermodynamics that the first law can be expressed as $dU = \dbar Q+ \dbar W = T dS + \sum_{n} F_{n} dx_{n}$ here $F_{n}$ is the generalized force and $x_{n}$ is the generalized co-ordinate corresponding to $F_n$.  Hence from this we can determine  the thermodynamic  entropy of the system and which is, 
\begin{equation}
dS =\frac{\dbar Q}{T}
\end{equation}
and holds only for the quasi-static processes.  
\section{Quantum Processes of the considered system}
This section describes the engine operation and the various processes involved in the cycle.  The considered model, composed of a two level system as a working substance, operates in a finite time duration.  The quantum Carnot like cycle, like its classical counterpart,  consists of four strokes: two isothermal strokes and two  Adiabatic strokes.  The system of spin half particles are assumed to be non-interacting (or a negligible interaction) and is subjected to external time dependent magnetic field.  When the system is attached (detached) from the reservoir of inverse temperature $\beta_h (\beta_c)$ and  if the process change quasi-statically from one state to another, then this process is an isothermal stroke and the path traced by this process is given by \cite{bassie}, 
\begin{equation}
\beta \Delta = ln \left( \frac{1-P^{e}}{P^{e}}\right)
\end{equation}
wherein the above, $\Delta = 2 \mu B$ denotes the energy gap between the ground and excited state of a spin half particle in the applied magnetic field, $B$, $\mu$ is the particle moment and $P^{e}$ is the probability of getting the particle in the excited state when the system is in contact with the reservoir.  \\

 For the considered model with the working medium of two level spin-half particles, the isothermal stroke itself comprises of two processes namely, 1) the adiabatic change and 2) the equilibration which takes place at  the constant occupation probability of the excited state and at constant magnetic field, respectively.   The process of adiabatic change occurs when the system is detached from the reservoir and vary the magnetic field for a very short time.  Here in this process no heat exchange takes place.  While equilibration occurs when the system is attached back to the reservoir by keeping the magnetic field constant.  Unlike the adiabatic change process, it is a relatively longer process and takes much time to reach the equilibrium.  Once equlilibrioum is reached, the system again perform the process of adiabatic change followed by the equilibration.  This alternate mode of operation continues till the system reaches the desired state.  Hence  the system's isothermal strokes  will consist of finite number of alternating modes of operation, viz., the adiabatic change and the equilibration to reach the destined state in finite time.  \\

\begin{figure}[H]
\centering
\includegraphics[width=\textwidth]{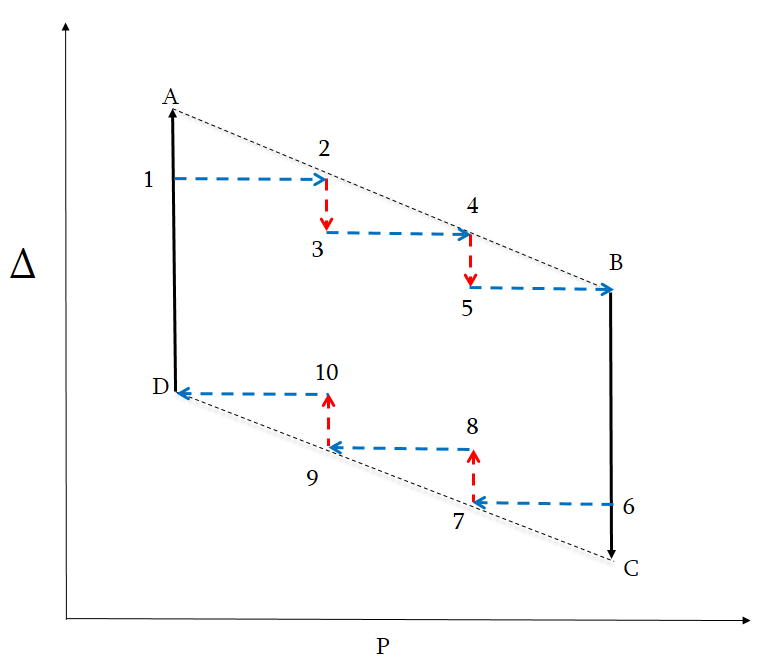}
\caption{Carnot like quantum heat engine model working between two reservoirs at different temperature under finite time process. Dashed line indicates the quasi static isothermal processes and the zig-zag segments represent the alternative adiabatic change and equilibration under finite time.} 
\label{f1}
\end{figure}

Figure \ref{f1} illustrates the quantum Carnot like cycle of the considered model 
\cite{bassie,kos}.  In the figure, $A\rightarrow B$ and $C \rightarrow D$ are the isothermal strokes.  Within the isothermal strokes, there are alternate processes of adiabatic change and equilibration which are represented by dotted red and blue arrows respectively. In the isothermal stroke $A\rightarrow B$, the successive processes occurring are $A \rightarrow 1$, $1 \rightarrow 2$, $2 \rightarrow 3$, $3 \rightarrow 4$, $4 \rightarrow 5$ and $5 \rightarrow B$.  Among these $A \rightarrow 1$ is an instantaneous process which occurs very fast.  The processes  $2 \rightarrow 3$ and $4 \rightarrow 5$ are the adiabatic change which occurs when the system is detached from the hot reservoir of inverse temperature, $\beta_h$ and the external field is changed infinitesimally to keep the system with constant probability.  Then the processes, $1 \rightarrow 2$, $3 \rightarrow 4$ and $5 \rightarrow B$ are the equlibration process which occurs when the magnetic field is kept constant and the system is attached back to the reservoir. This process continues until the system equlibrates with the reservoir of inverse temperature $\beta_h$.  
Similar to the isothermal stroke $A\rightarrow B$,  the isothermal stroke $C \rightarrow D$ also has successive processes which are $C \rightarrow 6$, $6 \rightarrow 7$, $7 \rightarrow 8$, $8 \rightarrow 9$, $9 \rightarrow 10$ and $10 \rightarrow D$.  Here, $7 \rightarrow 8$ and $9 \rightarrow 10$  are adiabatic change process, $6 \rightarrow 7$, $8\rightarrow 9$ and $10 \rightarrow D$ are the equilibration processes.  The successive section evaluates the heat and work done during the isothermal strokes.  \\

\subsection{Thermodynamic quantities associated with isothermal stroke}
\subsubsection{Isothermal stroke $A\rightarrow B$}

Step:1 $A\rightarrow 1$ (Detaching the sytem from reservoir):\\
The finite time process begins with the initial state $(P_{A}^{e}, \Delta_{A})$ and at that instant of time the system is kept in contact with the hot reservoir having inverse temperature $\beta_h$.  The system reaches thermal equilibrium with the reservoir.  It is then detached from the reservoir.  Now the energy level decreases from $\Delta _A$ to $\Delta_1$ abruptly during which the system can extract work.  The amount of work done by the system is given by equation~\ref{workdone}, 
\begin{equation}
W_{A\rightarrow 1} = P_{A}^{e} \left[ \Delta_{1}-\Delta_{A}\right]
\end{equation}
Step:2 $1\rightarrow 2$ (System is kept in contact with the reservoir):\\
In this process $1\rightarrow 2$, the system is allowed to equilibrate thermally by exchanging heat and the change of energy level ramains constant.  Hence the amount of heat abosrbed by the system is,
\begin{equation}
Q_{1\rightarrow 2} = \Delta_{1} \left[ P_{2}^{e}-P_{1}^{e}\right].  
\end{equation}
Step:3 $2\rightarrow 3$ (Detaching the sytem from reservoir):\\
Here again the system is kept in contact with the hot reservoir having inverse temperature $\beta_h$,  the energy level decreseas from $\Delta _2$ to $\Delta_3$ abruptly during which the system can extract work.  The amount of work done by the system during the process  $2\rightarrow 3$  is, 
\begin{equation}
W_{2\rightarrow 3} = P_{2}^{e} \left[ \Delta_{3}-\Delta_{2}\right]
\end{equation}
Step:4 $3\rightarrow 4$ (System is kept in contact with the reservoir):\\
In this process, the system equlibrates and due to the absorbed heat,  the population of the excited state 2, $(P_{2}^{e})$ increases.  Whereas the population of the state 1 decreases and is $(P_{1}^{e})$. The amount of heat abosrbed by the system during the process $3\rightarrow 4$ is,
\begin{equation}
Q_{3\rightarrow 4} = \Delta_{3} \left[ P_{4}^{e}-P_{3}^{e}\right].  
\end{equation}
Step:5 $4\rightarrow 5$ (Detaching the sytem from reservoir):\\
In a similar fashion as step 1 and step 3, the system can extract work during the process $4\rightarrow 5$ and the amount of work done by the system during this process  is, 
\begin{equation}
W_{4\rightarrow 5} = P_{4}^{e} \left[ \Delta_{5}-\Delta_{4}\right]
\end{equation}
Step:6 $5\rightarrow B$ (System is kept in contact with the reservoir):\\
This is the last process of the Isothermal stroke $A\rightarrow B$ (according to our consideration), in which the system equilibrates with the reservoir at inverse temperate $\beta_h$ maintaining the energy gap to be constant.  The amount of heat abosrbed by the system during the process $5\rightarrow B$ is,
\begin{equation}
Q_{5\rightarrow B} = \Delta_{B} \left[ P_{B}^{e}-P_{5}^{e}\right].  
\end{equation}
From the above equations, the net work done by the two level quantum Carnot like heat engine in the isothermal stroke  $A\rightarrow B$ and the net heat absorbed during the isothermal stroke $A\rightarrow B$. 
The network done in the isothermal stroke $A\rightarrow B$ is \cite{bassie},
\begin{equation}
W_{A \rightarrow B} = \Delta (P_{A}^{e}+ P_{2}^{e}+P_{4}^{e})
\label{workab}
\end{equation} 
where in the above equations $\Delta = \frac{\Delta_A -\Delta_B}{n}$, $n$ is the number of subdivisions and the total amount of heat aborbed during isothermal stroke $A\rightarrow B$ is \cite{bassie},
\begin{equation}
Q_{A \rightarrow B} = \Delta_{B} P_{B} ^{e} - \Delta_{A} P_{A} ^{e}+ \Delta \left[ P_{A}^{e}+P_{2}^{e}+P_{4}^{e}\right]
\end{equation} 
In the above stroke, the heat absorbed is positive and since the loses its internal energy, the workdone by the system is negative \cite{bassie}. 
\subsubsection{Isothermal stroke $C\rightarrow D$}
Similar to the isothermal stroke  $A\rightarrow B$, this branch also performs alternate steps of adiabatic change and equilibration which are listed below and the corresponding expression for heat and work are also given. \\
Step:1 $C\rightarrow 6$ (Detaching the sytem from reservoir):\\
In this mode, the population of the excited state is constant and change in energy level occurs between $\Delta_{C}$ to $\Delta_{6}$.  This is a very fast process and the state of the system at the begining of this step is $(P_{C}^{e}, \Delta_{C})$ followed by the detachment of the system from the reservoir with inverse temperature $\beta_c$.  The workdone during this sudden process is, 
\begin{equation}
W_{6\rightarrow C} = P_{C}^{e} \left[ \Delta_{6}-\Delta_{C}\right]
\end{equation}
Step:2 $6\rightarrow 7$ (System is kept in contact with the reservoir):\\
In this process $6\rightarrow 7$, the system is allowed to equilibrate thermally with the reservoir of inverse temperature $\beta_c$  by releasing heat and the change of energy level ramains constant.  Hence the amount of heat released by the system is,
\begin{equation}
Q_{6\rightarrow 7} = \Delta_{6} \left[ P_{7}^{e}-P_{6}^{e}\right].  
\end{equation}
Step:3 $7\rightarrow 8$ (Detaching the sytem from reservoir):\\
Here again the system is kept in contact with the cold reservoir having inverse temperature $\beta_c$,  the energy level increases from $\Delta _2$ to $\Delta_3$ abruptly during which the system can do work.  The amount of work done by the system during the process  $7\rightarrow 8$  is, 
\begin{equation}
W_{7\rightarrow 8} = P_{7}^{e} \left[ \Delta_{8}-\Delta_{7}\right]
\end{equation}
Step:4 $8\rightarrow 9$ (System is kept in contact with the reservoir):\\
In this process, the system equlibrates and due to the released heat, the population of the excited state 9, $(P_{9}^{e})$ decreases.  Whereas the population of the state 8 increases and is $(P_{8}^{e})$. The amount of heat released by the system during the process $8 \rightarrow 9$ is,
\begin{equation}
Q_{8\rightarrow 9} = \Delta_{8} \left[ P_{9}^{e}-P_{8}^{e}\right].  
\end{equation}
Step:5 $9\rightarrow 10$ (Detaching the system from reservoir):\\
The  work is done by the system during the process $9\rightarrow 10$ and the amount of work done  is given by, 
\begin{equation}
W_{9\rightarrow 10} = P_{9}^{e} \left[ \Delta_{10}-\Delta_{9}\right].
\end{equation}
Step:6 $10\rightarrow D$ (System is kept in contact with the reservoir):\\
This is the last process of the Isothermal stroke $C\rightarrow D$ (according to the considered model), in which the system equilibrates with the reservoir at inverse temperate $\beta_c$ maintaining the energy gap to be constant.  The amount of heat released by the system during the process $10\rightarrow D$ is,
\begin{equation}
Q_{10\rightarrow D} = \Delta_{D} \left[ P_{D}^{e}-P_{10}^{e}\right].  
\end{equation}
The total amount of work done by the system during the isothermal stroke ${C\rightarrow D}$ is,
\begin{equation}
W_{C \rightarrow D} = \Delta^{'} (P_{7}^{e}+ P_{9}^{e}+P_{C}^{e})
\label{workcd}
\end{equation} 
and the total amount of heat released during isothermal stroke $C\rightarrow D$ is \cite{bassie},
\begin{equation}
Q_{C \rightarrow D} = \Delta_{D} P_{D} ^{e} - \Delta_{C} P_{C} ^{e}+ \Delta^{'} \left[ P_{C}^{e}+P_{7}^{e}+P_{9}^{e}\right]
\end{equation} 
where in the above equations $\Delta^{'} = \frac{\Delta_D -\Delta_C}{n}$, $n$ is the number of subdivisions.
\subsection{Thermodynamic quantities associated with adiabatic stroke}
The stroke $B \rightarrow C$ is one of the adiabatic stroke which begins when the system reaches the satate $B$, it is then detached from the hot reservoir $\beta_h$ and connected to the cold reservoir $\beta_c$.  Here the energy gap changes when the system is changed from one reservoir to another, whereas the population of the spin half particles in the two states $B$ and $C$ are same i.e. $P_{B} ^{e}=P_{C} ^{e}$.  During the adiabatic stroke there is no heat exchange, and hence $Q=0$.  The amount of work done by the system is given by \cite{bassie},
\begin{equation}
W_{B \rightarrow C} =  P_{B} ^{e} \left[ \Delta_{C} - \Delta_{B}\right].  
\end{equation} 
Similarly, the other adiabatic stroke occurs when the system reaches the state $D$.  It is then detached from cold reservoir $\beta_c$ to the hot reservoir $\beta_h$.  In this stroke, the energy gap shifts from $\Delta_D$ to $\Delta_A$ and the populations of the states remain unchanged i.e., $P_{D} ^{e}=P_{A} ^{e}$.  The amount of work done by the system during this adiabatic stroke is, 
\begin{equation}
W_{D \rightarrow A} =  P_{A} ^{e} \left[ \Delta_{A} - \Delta_{D}\right].  
\end{equation} 
The net amount of heat and work done by the system per cycle for the system of spinhalf particles is given by, 
\begin{equation}
Q_{net} =  \Delta_{B} P_{B} ^{e} - \Delta_{A} P_{A} ^{e}+ \Delta\left[P_{A} ^{e}+P_{2} ^{e}+P_{4} ^{e}\right] +\Delta_{D} P_{D} ^{e} - \Delta_{C} P_{C} ^{e}+ \Delta\left[P_{C} ^{e}+P_{7} ^{e}+P_{9} ^{e}\right] 
\end{equation} 
and 
\begin{equation}
W_{net} =  (\Delta_{C}-\Delta_{B}) P_{B} ^{e}  \left[1+\frac{(\Delta_{A}-\Delta_{D})P_{A} ^{e}}{(\Delta_{C}-\Delta_{B})P_{B} ^{e}}\right ] + \Delta^{'} P_{C} ^{e} - \Delta P_{A} ^{e}+\left(\Delta^{'}-\Delta \right)  \left( P_{2} ^{e}+P_{4} ^{e}\right).
\end{equation} 
And hence from the above results, the efficiency of the two level quantum heat engine of spin half particles is then given by \cite{bassie}, 
\begin{equation}
\eta = \frac{W_{net}}{Q_{A\rightarrow B}}=\frac{(\Delta_{B}-\Delta_{C}) P_{B} ^{e}  \left[1-\frac{(\Delta_{A}-\Delta_{D})P_{A} ^{e}}{(\Delta_{C}-\Delta_{B})P_{B} ^{e}}\right ]}{\Delta_{B} P_{B} ^{e} - \Delta_{A} P_{A} ^{e}+ \Delta  \left[P_{A} ^{e}+P_{2} ^{e}+P_{4} ^{e}\right]}  -  \frac{\Delta^{'} P_{C} ^{e} + \Delta P_{A} ^{e}-\left(\Delta^{'}-\Delta \right)  \left( P_{2} ^{e}+P_{4} ^{e}\right)}{\Delta_{B} P_{B} ^{e} - \Delta_{A} P_{A} ^{e}+ \Delta\left[P_{A} ^{e}+P_{2} ^{e}+P_{4} ^{e}\right]}.
\label{efficiency}
\end{equation}

\section{Efficiency at maximum $\dot{\Omega}$ figure of merit}
In this section, we derive the expression for efficiency at maximum  $\dot{\Omega}$ figure of merit, which provides the compromise between useful energy and the lost energy.  We can also compare the efficiency at maximum  $\dot{\Omega}$ figure of merit with the Curzon-Ahlborn efficiency.  The $\dot{\Omega}$ figure of merit is defined through,
\begin{equation}
\dot{\Omega} = \frac{(2 \eta - \eta_{max})P} {\eta}.
\label{omeg}
\end{equation} 
As we know, $\eta_{max}$ is the maximum efficiency which is nothing but the Carnot efficiency $\eta_C$, $\eta =\frac{W_{net}}{Q_{A\rightarrow B}}$ and the power $P =\frac{W_{net}}{t}$, where $t$ is the time period. With the inclusion of $P$ and $\eta$ in the equation~\ref{omeg}, we have,
\begin{equation}
\dot{\Omega} = \frac{(2 \eta -\eta_C) Q_{A\rightarrow B}}{t}. 
\end{equation} 
On substituting the expressions for $Q_{A\rightarrow B}$ in the above equation, one can get, 
\begin{eqnarray}
\dot{\Omega} &=& \frac{1}{t}\big[\Delta_{B} P_{B} ^{e} - \Delta_{A} P_{A} ^{e}+ \Delta \left[ P_{A}^{e}+P_{2}^{e}+P_{4}^{e}\right]\big]  \times \\ \nonumber
& &\left[2\left(\frac{(\Delta_{B}-\Delta_{C}) P_{B} ^{e}  \left[1-\frac{(\Delta_{A}-\Delta_{D})P_{A} ^{e}}{(\Delta_{C}-\Delta_{B})P_{B} ^{e}}\right] -\left(\Delta^{'} P_{C} ^{e} + \Delta P_{A} ^{e}-\left(\Delta^{'}-\Delta \right)  \left( P_{2} ^{e}+P_{4} ^{e}\right) \right)}{\Delta_{B} P_{B} ^{e} - \Delta_{A} P_{A} ^{e}+ \Delta  \left[P_{A} ^{e}+P_{2} ^{e}+P_{4} ^{e}\right]}\right)-\eta_{C}\right]
\label{omeg2}
\end{eqnarray}

As mentioned previously while defining $\Delta=\frac{(\Delta_{A}-\Delta_{B})}{n}$ and $\Delta^{'}=\frac{(\Delta_{A}-\Delta_{B})}{n}$,  n is the number of divisions. This number of subdivision is actually proportional to $t$, i.e., $n \propto t$ \cite{bassie}. Then the above equation  can be rewritten as, 
\begin{eqnarray}
\dot{\Omega}&=&\frac{1}{t} \left[\Delta_{B} P_{B} ^{e} - \Delta_{A} P_{A} ^{e}+ \frac{(\Delta_{A}-\Delta_{B}) \left[ P_{A}^{e}+P_{2}^{e}+P_{4}^{e}\right]}{t}\right] \times \\ \nonumber
& &\left[2\left(\frac{\left[(\Delta_{B}-\Delta_{C}) P_{B}^{e}\left[1-\frac{(\Delta_{A}-\Delta_{D})P_{A} ^{e}}{(\Delta_{C}-\Delta_{B})P_{B} ^{e}} \right]\right]-\left[\frac{(\Delta_{D}-\Delta_{C})P_{C} ^{e}-(\Delta_{A}-\Delta_{B})P_{A} ^{e}+(\Delta_{D}-\Delta_{C}-\Delta_{A}+\Delta_{B})(P_{2} ^{e}+P_{4} ^{e})}{t} \right]}{\Delta_{B}P_{B} ^{e}-\Delta_{A}P_{A} ^{e}+\frac{(\Delta_{A}-\Delta_{B})(P_{A} ^{e}+P_{2} ^{e}+P_{4} ^{e})}{t}}\right)-\eta_{C}\right].             
\label{omegfinal}
\end{eqnarray}
Optimizing the $\dot{\Omega}$ figure of merit with respect to time $t$ $(\frac{\partial \dot{\Omega}}{\partial t} =0)$ gives the values of $\widetilde{t}$ at which $\dot{\Omega}$ figure of merit is maximum. The value of $\widetilde{t}$ is given below:
\begin{equation}
\widetilde{t} = \frac{2\left(2\left[(\Delta_{D}-\Delta_{C}) P_{C} ^{e}-(\Delta_{A}-\Delta_{B}) P_{A} ^{e}+(\Delta_{D}-\Delta_{C}-\Delta_{A}+\Delta_{B})( P_{2} ^{e}+ P_{4} ^{e})\right]+\eta_{C}\Phi
\right)}{(\Delta_{B}-\Delta_{C}) P_{B} ^{e}\left[1-\frac{(\Delta_{A}-\Delta_{D}) P_{A} ^{e}}{(\Delta_{B}-\Delta_{C})P_{B} ^{e}}\right]},
\label{time2}
\end{equation}
where $\Phi=(\Delta_{A}-\Delta_{B})( P_{A} ^{e}+ P_{2} ^{e}+ P_{4} ^{e})$.
Using the equation~\ref{time2} in equation~\ref{efficiency}, we can obtain the efficiency at maximum $\dot{\Omega}$ figure of merit and is given by, 
\begin{equation}
\eta_{\dot{\Omega}_{max}} = \eta_C \left\{\frac{3\left[(\Delta_{D} -\Delta_{C})P^{e}_{C}-(\Delta_{A} -\Delta_{B})P^{e}_{A}+(\Delta_{D} -\Delta_{C}-\Delta_{A}+\Delta_{B})(P^{e}_{2}+P^{e}_{4}) \right]+2\eta_{C} \Phi}{4\left[(\Delta_{D} -\Delta_{C})P^{e}_{C}-(\Delta_{A} -\Delta_{B})P^{e}_{A}+(\Delta_{D} -\Delta_{C}-\Delta_{A}+\Delta_{B})(P^{e}_{2}+P^{e}_{4}) \right]+3\eta_{C}\Phi}   \right\}.
\end{equation} 
After rearranging the above equation, the efficiency at maximum $\dot{\Omega}$ figure of merit of a two level quantum heat engine becomes, 
\begin{equation}
\eta_{\dot{\Omega}_{max}} = \eta_{C} \left[\frac{2 \eta_{C} + 3\frac{\Gamma}{\Phi}}{3 \eta_{C} + 4\frac{\Gamma}{\Phi}}\right],
\label{effiomeg1}
\end{equation}
where $\Gamma =\Pi-\Phi$ and $\Pi=(\Delta_{D} -\Delta_{C})(P^{e}_{C}+P^{e}_{2}+P^{e}_{4})$.

It can be observed from the above that the value of efficiency at maximum $\dot{\Omega}$ figure of merit depends on the population probabilities and the energy gap between the excited and ground states, which is given by the ratio $\frac{\Gamma}{\Phi}$ (equation~\ref{effiomeg1}).  For the positive values of $\frac{\Gamma}{\Phi}$, the generalized extreme bounds of the efficiency at maximum $\dot{\Omega}$ figure of merit are obtained from equation~\ref{effiomeg1} as, 
\begin{equation}
\frac{2}{3} \eta_{C} \equiv \eta_{\dot{\Omega}_{max}}^{-}\leq\eta_{\dot{\Omega}_{max}}\leq\eta_{\dot{\Omega}_{max}}^{+}\equiv \frac{3}{4}\eta_{C}
\label{bounds}
\end{equation}
These lower and upper bounds of the the efficiency at maximum $\dot{\Omega}$ figure of merit are achieved when $\frac{\Gamma}{\Phi}\longrightarrow 0$ and $\frac{\Gamma}{\Phi}\longrightarrow \infty$, respectively. 
This can be obtained for the  positive extreme values of 
$\frac{\Gamma}{\Phi}$ when $\Phi \longrightarrow  \Pi$ and $\Phi \longrightarrow 0$, respectively.

 \begin{figure}[H]
\centering
\includegraphics[width=\textwidth]{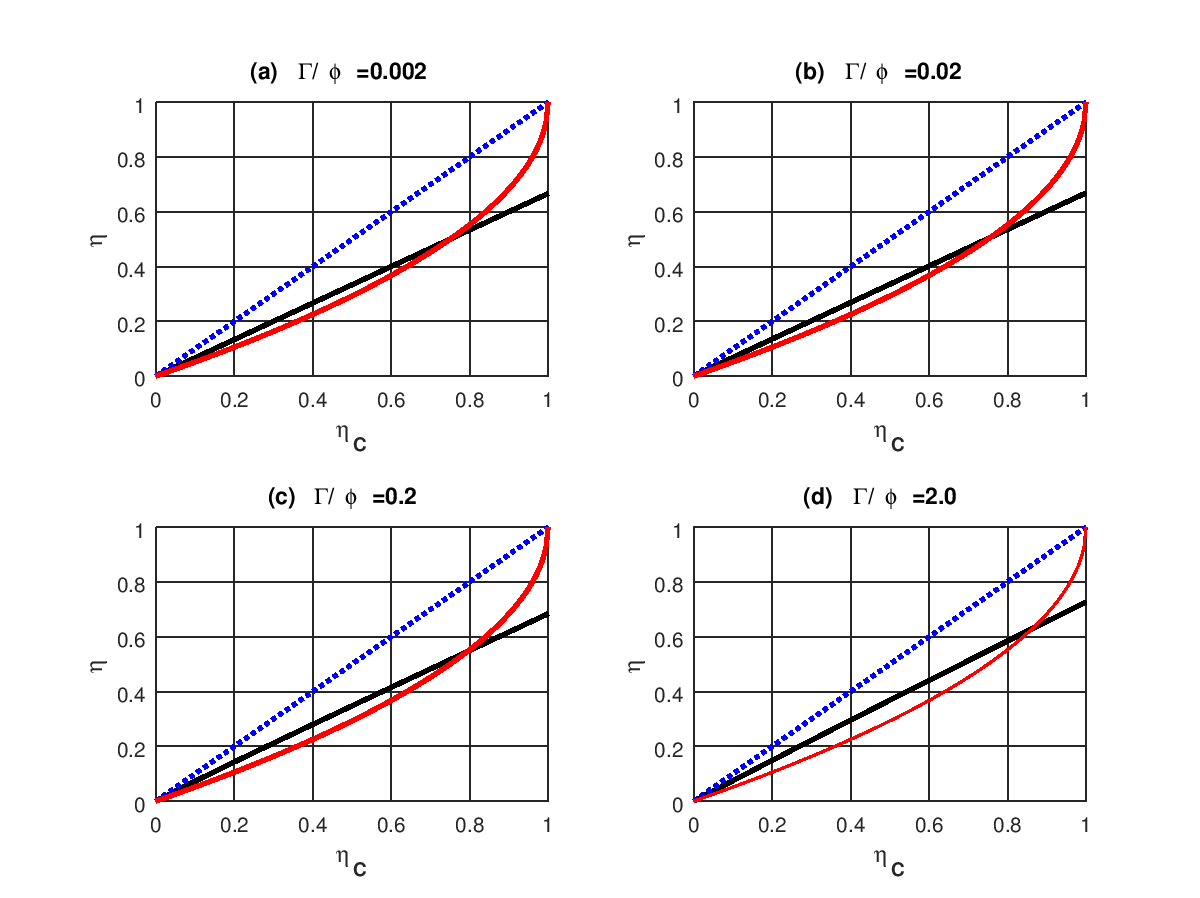}
\caption{The plot of efficiency at maximum $\dot{\Omega}$ figure of merit (solid black line) for various values of $\Gamma/\phi$ together with Carnot efficiency (dotted blue line) and Curzon-Ahlborn efficiency (solid red line). } 
\label{f2}
\end{figure}

The efficiency at maximum $\dot{\Omega}$ figure  of merit thus obtained is plotted in Figure \ref{f2} as the functions of the Carnot efficiency for the positive values of $\frac{\Gamma}{\Phi}$.  From the figure it is observed that, the efficiency at maximum $\dot{\Omega}$ figure of merit lies below  $\eta_{CA}$ for the  higher values of $\eta_{C}$ and it becomes higher than the Curzon-Ahlborn efficiency  when $\eta_{C}$ decreases for different values of $\frac{\Gamma}{\Phi}$.
We also observed that there is a cross over of efficiency at maximum $\dot{\Omega}$ figure of merit
with $\eta_{CA}$ and the value  at which this cross over occurs decreases with  decrease in  $\frac{\Gamma}{\Phi}$. It is also found that the value of the cross over point gets shifted towards $\eta_{C}$ around $0.75$ when $\frac{\Gamma}{\Phi} \rightarrow 0$.

\begin{figure}[H]
\centering
\includegraphics[width=\textwidth]{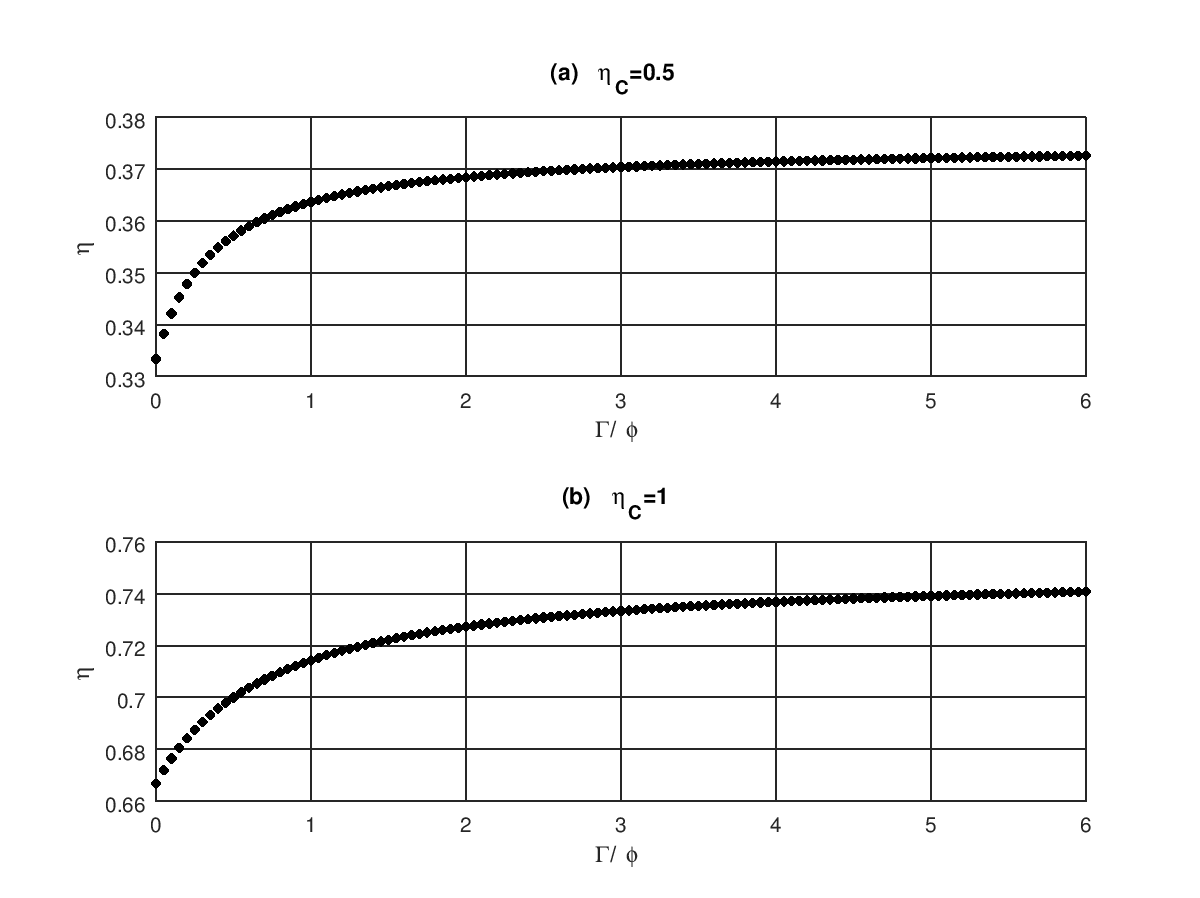}
\caption{The plot of efficiency at maximum $\dot{\Omega}$ figure of merit for various positive values of $\Gamma/\phi$
for two different values of  Carnot efficiency.} 
\label{f3}
\end{figure}

In addition to this, there is an interesting fact to observe in the efficiency at maximum $\dot{\Omega}$ figure of merit. For two different values of $\eta_{C}$, the plot for efficiency at maximum $\dot{\Omega}$ figure of merit versus $\frac{\Gamma}{\Phi}$ is shown in Figure \ref{f3}. From the extreme bounds of the efficiency at maximum $\dot{\Omega}$ figure of merit ~Equation \ref{bounds}, we can observe that, the lower bound is the efficiency corresponding to the efficiency at maximum efficient power and the upper bound is the one corresponding to the efficiency at maximum $\dot{\Omega}$ figure of merit. This implies, the present model of quantum Carnot like heat engine can deliver the maximum efficiency whose extreme limits are the lower bound of efficiency obtained by maximizing 
the efficiency at two different figure of merits viz., efficient power and  $\dot{\Omega}$ figure of merit \cite{nila}. 

\begin{figure}[H]
\centering
\includegraphics[width=\textwidth]{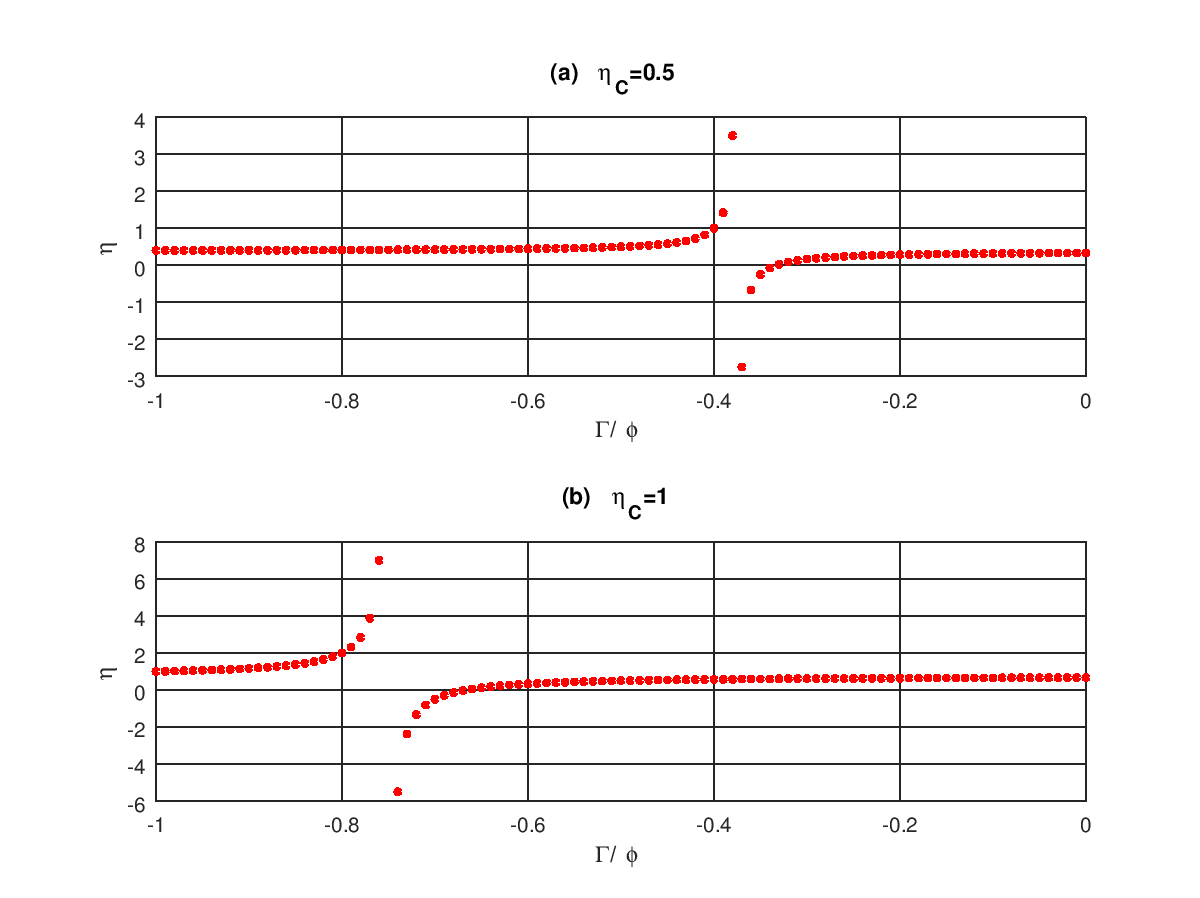}
\caption{The plot of efficiency at maximum $\dot{\Omega}$ figure of merit for various negative values of $\Gamma/\phi$ for two different values of  Carnot efficiency.} 
\label{f4}
\end{figure}

In the above analysis, we consider only the positive values of $\frac{\Gamma}{\Phi}$, however, one can not simply ignore the possibility of negative values of $\frac{\Gamma}{\Phi}$, since this provide the  physical situation that there can be more  number of spin half particles in higher energy states than the lower energy states. In such  circumstances, 
the positive value of $\eta_{\dot{\Omega}_{max}}$ in equation~\ref{effiomeg1} imposes a restriction on the negative values of $\frac{\Gamma}{\Phi}$.  
For two different values of $\eta_{C}$, the plot for efficiency at maximum $\dot{\Omega}$ figure of merit versus negative values $\frac{\Gamma}{\Phi}$ is shown in Figure \ref{f4}. From this figure it is observed an anomalous behavior in the region $-(3/4) \eta_{C}< \frac{\Gamma}{\Phi}<-(2/3)\eta_C$. In particular, $\eta_{\dot{\Omega}_{max}}$ reaches negative values when $\frac{\Gamma}{\Phi}=-(2/3)\eta_C$ and goes on decreases. This behavior persists till  $\frac{\Gamma}{\Phi}$ reaches $-(3/4)\eta_C$ and one can observe a singularity in
efficiency at maximum $\dot{\Omega}$ figure of merit in this region. Once $\frac{\Gamma}{\Phi}$ crosses $-(3/4)\eta_C$, the efficiency at $\dot{\Omega}$ figure of merit exceeds all the bounds and reaches the Carnot efficiency when we move far away from $-(3/4)\eta_C$ as shown in Figure \ref{f4}.   This trend of the efficiency at maximum $\dot{\Omega}$ figure of merit crossing all the bounds and even reaching 
$\eta_{C}$ in the finite time duration indicates the coherent nature of collective population occupancy behavior of spin half particles which play a important role of quantum heat engine in the negative region of $\frac{\Gamma}{\Phi}$. Even though the  particles are assumed to be non-interacting, the thermal interaction in the negative region of $\frac{\Gamma}{\Phi}$ sufficient to provides the considerable collective effect of the particles in the population levels. This collective effect \cite{coll} provides a possibility to construct  efficient quantum heat engines with working substance as the collection of spin half particles \cite{exp}.

\section{Conclusion}
In this work, we have considered a  quantum Carnot like heat engine comprising two level spin half system as a working substance operates in a finite time duration.  We derived the efficiency of the quantum Carnot like heat engine  at maximum $\dot{\Omega}$ figure of merit. The cycles are considered to be analogous to the classical counterpart except that the isothermal stroke itself composed of alternative process of adiabatic change and equilibration. The extreme bounds obtained at maximum $\dot{\Omega}$ figure of merit corresponds to the lower bound of efficiency obtained at maximum efficient power and maximum $\dot{\Omega}$ figure of merit. It is also found that engine efficiency at maximum $\dot{\Omega}$ figure of merit exceeds the above bounds and also approaches Carnot efficiency in certain population regime. Recent study of thermodynamics of a minimal collective heat engine \cite{coll} showed that with the proper design of interaction, we can achieve the efficiency (which is optimized at maximum power) much more than the known bounds and even approaches the Carnot efficiency $\eta_C$.  Our result indicates that the collection of spin half particles as working substance in quantum heat engine can be a better choice to design minimal collective heat engine in the quantum regime which provides superior engine performance with high efficiency.

\end{document}